\documentclass[twocolumn,aps,showpacs,nofootinbib,epsfig]{revtex4}
\usepackage{graphicx}
\usepackage{dcolumn}
\usepackage{bm}
\usepackage{amssymb}
\usepackage{amsmath}
\usepackage{epsfig}
\usepackage{ucs} 
\usepackage{amsmath} 
\usepackage{array}
\usepackage{float}
\usepackage{color}
\usepackage[usenames,dvipsnames]{xcolor}
\usepackage{epstopdf}
\usepackage{natbib}
\usepackage{lineno}
\usepackage{slashed}
\usepackage{color}
\usepackage{caption}
\usepackage{hyperref}
\definecolor{red}{rgb}{1.0,0.0,0.0}
\definecolor{blue}{rgb}{0.0,0.0,1}

\newcommand{\n}{\mathbf}
\newcommand{\bea}{\begin{eqnarray}}
\newcommand{\eea}{\end{eqnarray}}
\newcommand{\nn}{\nonumber}
\newcommand{\etal}{\textit{et al.}}

\newcommand{\wc}{\omega_c}

\usepackage{soul}
\begin{document}
%
\title{Super-statistical description of thermo-magnetic properties of a system of 2D GaAs quantum dots with gaussian confinement and Rashba spin-orbit interaction}

\author{Jorge David Casta\~no-Yepes$^1$\footnote{Corresponding author.\\
E-mail address: jorgecastanoy@gmail.com (J.D. Castaño-Yepes).} and D. A. Amor-Quiroz$^{2}$}
\address{
  $^1$Instituto de Ciencias
  Nucleares, Universidad Nacional Aut\'onoma de M\'exico, Apartado
  Postal 70-543, M\'exico Distrito Federal 04510,
  Mexico.\\
  $^2$CPHT, CNRS, Ecole Polytechnique, Institut Polytechnique de Paris, Route de Saclay, 91128 PALAISEAU.}
%
\begin{abstract}
We examine the effect of non-equilibrium processes modeled by the introduction of a generalized Boltzmann factor on the thermal and magnetic properties of an array of two-dimensional GaAs quantum dots in the presence of an external uniform and constant magnetic field. The model consists of a single-electron subject to a confining Gaussian potential with a spin-orbit interaction in the Rashba approach. We compute the specific heat and the magnetic susceptibility within the formalism of $\chi^2$-superstatistics from the exact solution of the Schr\"odinger equation. Furthermore, an analytic solution for the partition function allows a study of the impact of the number of subsystems on the superstatistical corrections and confirms that the ordinary thermo-magnetic properties are recovered whenever the thermal distribution can be approximated by a Dirac delta. Also, we found a progressive disappearance of the Schottky anomaly with decreasing number of subsystems, while the specific heat ceases to be a monotonically increasing function with respect to the average temperature when the $\chi^2$-distribution is spread over a large range of temperatures.
Remarkably, the introduction of fluctuations in the temperature is found to suppress the paramagnetic phase transition that would otherwise appear at low temperatures. Finally, we emphasize that an appropriate construction of the definition of physical observables is crucial for obtaining a correct description of the physics derived from a non-extensive construction of the entropy.
\end{abstract}

\maketitle
\section{Introduction}

Semiconductor quantum dots (QDs), also called artificial atoms, consist of charge carriers confined in all directions resulting in a discrete energy spectrum similar to atoms in nature. Such systems are very attractive both from an experimental and a theoretical point of view as they possess a wide range of technological applications in quantum information and photoelectronics \cite{QuantumInfo1,QuantumInfo2,QuantumInfo3,QuantumInfo4}. This is due to the possibility of controlling several characteristics of such systems. For example, it is nowadays possible to vary with high precision the number of electrons composing a QD \cite{mesoscopic1, mesoscopic2, mesoscopic3, mesoscopic-self-citation}, the size \cite{tunable-size,tunable-size2}, the shape \cite{tunable-shape} and their composition~\cite{composition}. This  leads to direct light-emission technologies such as diodes~\cite{diodes1,diodes2}, displays~\cite{displays}, luminescent solar concentrators~\cite{solarconcentrator1,solarconcentrator2}, lasers~\cite{lasers1,lasers2}, solar cells~\cite{solarcells}, among others~\cite{reviewQDs}.

In particular, spin-related phenomena in QDs has been studied in extension in the last decades as they are crucial in the semiconductor technology called spintronics \cite{Wolf,spintronicsreview,pogressinspintronics}. For instance, the spin-orbit (SO) coupling mechanisms in semiconductors~\cite{Halperin,Voskoboynikov,Denis,Destefani,Marian,Voskoboynikov2} provide a basis for device applications and a source of interesting physics such as the spin transistor~\cite{Datta, spintransistoroverwiev}.  The Rashba effect is of particular interest as it provides a SO coupling whose tunability allows SO effects to occur in QDs with few electrons \cite{Governale}. In fact, some theoretical studies of the impact of the Rashba-SOI on the optical properties of a disk-like QD in the presence of an external magnetic field have been carried out within the framework of the density matrix approach \cite{Hosseinpour}. Notably, a spin dependence of the spatial wave functions is typical in systems involving a Rashba QD in the presence of an external magnetic field, which can affect the thermo-magnetic and optical properties of a QD~\cite{SchottkyAnomalyGumber, packman}.

From  a  theoretical  point  of  view, a correct description of the confining  potential  is  the  key  to  find  new  phenomena in the QDs dynamics. By choosing a suitable function of the confining potential and interactions, it is possible in some cases to analytically find the energy spectrum of the system, which allows the theoretical exploration of the thermal, magnetic and optical properties of QD's. It is well established that a harmonic potential is a good approximation which reproduces the main characteristics of such systems~\cite{Castano, Sanjeev2}, but it has been demonstrated that the confinement is rather anharmonic and has a finite depth which has been simulated by several authors using a Gaussian potential model~\cite{Adamowski,GaussianPotential1,GaussianPotential2,GaussianPotential3,GaussianPotential4,GaussianPotential5,GaussianPotential6,packman}. These works are developed in the context of the Boltzmann-Gibbs statistics, i.e., the QD under study represents the whole system which is in thermal equilibrium with an external bath. Therefore, it is possible to apply the canonical partition function formalism for computing the thermal properties from the spectrum of the confined electrons.

A more general thermal description of the system can take into account the fluctuations in the temperature or any of its intensive observables and thus, the usual Boltzmann-Gibbs formulation cannot be used. In such scenarios, it is necessary to achieve the correct formalism to describe situations out of the thermodynamic equilibrium. Superstatistics (SE) is one of the most attractive tools to describe the non-equilibrium dynamics of complex systems~\cite{Sobhani1, Sobhani2, Sargolzaeipor1, Sargolzaeipor2, Sargolzaeipor3, Chung, Hassanabadi, Sargolzaeipor4}, given its applications in several fields where the dynamic exhibit inhomogeneous spatiotemporal properties and making necessary an extension of the usual statistical methods. Beck and Cohen~\cite{BeckCohen1,BeckCohen2} introduced the SE formalism as a generalization of the Boltzmann-Gibbs factor $e^{-\beta\hat{H}}$ through the assumption of the existence of fluctuations in an intensive parameter $\tilde{\beta}$. This parameter can be identified as a local or average inverse temperature, effective friction constant, a changing mass parameter,  volatility in finance, changing noise, average chemical potential or some quantity which fluctuates on a much larger time scale than the typical relaxation time of the local dynamics. The applications of SE cover several fields and topics: hydrodynamical turbulence models~\cite{Turbulence1, Turbulence2, Turbulence3, Turbulence4, Turbulence5, Turbulence6, Turbulence7, Turbulence8, Turbulence9, Turbulence10}, cosmic ray statistics~\cite{CohenCosmicRays}, Quantum Chromodynamics phase diagram~\cite{AyalaCEP}, heavy-ion collisions~\cite{wilk1, wilk2, wilk3, Rybczynski, wong, wilk4, Bialas1, Bialas2, Rozynek, Tripathy1, Grigoryan, Khuntia, Bhattacharyya, Tripathy2,Ishihara, wilk5}, generalized entropies and generalized Newton's law~\cite{Obregon1, Merino, Obregon2}, Networks~\cite{Abe, Hasegawa}, momentum spectra of hadronic particles produced in $e^{-}e^{+}$ annihilation experiments~\cite{Bediaga}, among others. The applicability of SE has also recently reached the physics of QDs, for example, it has been used for describing the thermodynamic properties of a set of QDs as a function of the thermal distribution among them~\cite{SSQD}. 

The search for possible out-of-equilibrium modifications to the response functions of a QD, such as the specific heat or the magnetization, is attractive in terms of possible applications and general understanding of the dynamics of such systems. It is well-known that response functions are susceptible to changes in shape, interactions, and material constants of QDs~\cite{GaussianPotential5,Castano, packman,heatcapacityGaAsQD,Abbarchi,twodonors}. Therefore, one can expect deviations from the typical behavior within a SE prescription.

In this paper, we focus our attention on the effects of the size of a system consisting of $N$- two dimensional single-electron GaAs QDs with Gaussian confinement and Rashba Spin-Orbit coupling. Each subsystem is assumed to be in local thermal equilibrium so that the usual statistical framework well describes its individual thermodynamic properties while thermal fluctuations are allowed between the composing subsystems. Such out-of-equilibrium scenario is described within the $\chi^2$-SE formalism.  All the system is in the presence of an external and constant magnetic field. The paper is organized as follows:
In Section~\ref{sec:formalism} we give a description of the model by first providing a solution to the single particle Schr\"odinger equation (Subsection~\ref{subsec:QD}). Afterwards we provide a review of the SE construction of the partition function in Subsection~\ref{subsec:SE}, followed by the application of such formalism for computing the thermomagnetic properties of the $N$ subsystems of single-electron 2D-QDs in Subsection~\ref{Subsec:Z}. In Section~\ref{sec:results} we provide the numerical results along with the physical interpretation of the data. The results are summarized in Sec.~\ref{sec:conclusions}.

\section{Theoretical model}\label{sec:formalism}

\subsection{Energy spectrum of the single quantum subsystem}
\label{subsec:QD}
In the present model, we study a system made of $N$  single-electron 2D-QDs in the presence of an external uniform and constant magnetic field. Each QD can be regarded as a different subsystem with local thermal equilibrium but obeying a $\chi^2$-distribution of the temperatures amongst them. This is a good approximation if the QDs are embedded in a low thermal conductivity material. The single-particle Hamiltonian of the QDs in the presence of an external magnetic field with both Zeeman and Rashba SO terms is given by
\bea
\hat{H}=\frac{1}{2m^*}\left(\n{p}-\frac{q}{c}\n{A}\right)^2+\hat{H}_{\text{G}}+\hat{H}_{\text{R}}+\frac{1}{2}\mu_Bg^*\hat{\n{S}}\cdot \n{B}.
\label{Hamiltoneano}
\eea

The first term in Eq.~(\ref{Hamiltoneano}) refers to the minimal coupling between a particle with charge $q$ and a vector field propagating with the speed of light $c$. By ignoring the effects produced by the conduction band electrons, the effective electron mass $m^*$ is assumed to be constant and thus the non-parabolicity of the conduction band is neglected~\cite{conductionband1,conductionband2,conductionband3}. Furthermore, in the coordinate representation, the vector potential $\n{A}$ is expressed in the symmetric gauge $\n{A}=\frac{B}{2}(-y,x,0)$ which in a polar-coordinate system has the form
\bea
\n{A}(r)=\frac{Br}{2}\n{e}_{\theta}.
\eea

The second term in Eq.~(\ref{Hamiltoneano}) is the confining Gaussian potential $\hat{H}_{\text{G}}$~\cite{GaussianPotential1,GaussianPotential2,packman} with:
\bea
\hat{H}_{\text{G}}&\doteq &-V_0 e^{-r^2/2R^2},\nn\\
&\approx&\frac{1}{2}m^*\omega^2r^2-V_0.
\eea
Here $V_0$ and $R$ define the depth and the range of the potential respectively, while the effective confining frequency $\omega$ is given by
\bea
\omega^2=2V_0 \left(\frac{\sqrt{\omega_c^2+\omega_h^2}}{\hbar+2m^*\sqrt{\omega_c^2+\omega_h^2}R}\right),
\label{GaussianPotential}
\eea 
where $\omega_c=qB/m^*$,  $\omega_h^2=V_0/m^* R^2$ and $\hbar$ is the Planck constant.

The third term in Eq.~(\ref{Hamiltoneano}) 
corresponds to the Rashba spin-orbit coupling $\hat{H}_{\text{R}}$ and is given by the general expression~\cite{SOIBerry,SO1}
\bea
\hat{H}_{\text{R}}&\doteq&\frac{\gamma}{\hbar}\vec{\sigma}\cdot\left[\nabla V\times\left(\n{p}-\frac{q}{c}\n{A}\right)\right]_z\nn\\
&=&\gamma\sigma_z\frac{d\hat{H}_{\text{G}}}{dr}\left[-i\left(\frac{1}{r}\right)\frac{\partial}{\partial\theta}+\frac{q}{2\hbar}Br\right],
\eea
where the Rashba coupling constant is denoted by $\gamma$ and $\vec{\sigma}=(\sigma_x,\sigma_y,\sigma_z)$ is the Pauli matrices vector, so that
\bea
\hat{H}_{\text{R}}\doteq\frac{1}{2}\gamma m^*\omega^2\left(s\frac{m^*\omega_c}{\hbar}r^2-2i\sigma_z\frac{\partial}{\partial\theta}\right)
\eea
with $s=\pm1$ referring to the spin projection. 

The last term in Eq.~(\ref{Hamiltoneano}) is the Zeeman coupling of the external magnetic field ${\mathbf{B}}$ with the electron spin ${\mathbf{\hat{S}}}$, where $\mu_B$ is the Bohr magneton and $g^*$ is the effective Land\'e factor of the electron. 

With all of the above, the Hamiltonian of the system becomes
\bea
\hat{H}&\doteq&-\frac{\hbar^2}{2m^*}\left[\frac{1}{r}\frac{\partial}{\partial r}\left(r\frac{\partial}{\partial r}\right)+\frac{1}{r^2}\frac{\partial^2}{\partial\theta^2}\right]+\frac{\wc}{2}\hat{L}_z\nn\\
&+&\frac{1}{2}m^*\Omega^2_s\,r^2-i\sigma_z\gamma m^*\,\omega^2\frac{\partial}{\partial\theta}+\frac{1}{2}\mu_Bg^*\hat{\n{S}}\cdot \n{B}-V_0,\nn\\
\label{Hamiltonian}
\eea
where $\hat{L}_z$ is the orbital angular momentum operator in 2D and we have defined 
\bea
\Omega^2_s=\left(1+s\gamma\frac{m^*\wc}{\hbar}\right)\omega^2+\left(\frac{\omega_c}{2}\right)^2.
\label{Omega_S}
\eea
The eigenvalues of the Hamiltonian of Eq. (\ref{Hamiltonian}) have been studied in a recent work~\cite{packman} and are given by
\bea
E_{nls}&=&\hbar\Omega_s \left(2n+|l|+1\right)-\frac{1}{2}\hbar\wc l\nn\\
&+&\left(\gamma m^*\omega^2l+\frac{1}{4}g^*\hbar\omega_c\right)s-V_0.
\label{energyspectrum}
\eea
The last equation allows us to compute the canonical partition function $\mathcal{Z}_0$ in an analytical form.

\subsection{Superstatistical description of the system}\label{subsec:SE}

The superstatistics was well described by Beck and Cohen~\cite{BeckCohen1} as an extension of the usual statistical description of a systems that has not yet reached the full equilibrium. The fluctuations are encoded in the intensive parameter $\tilde{\beta}$ in such a way that the whole system can be divided in subsystems where $\tilde{\beta}$ is approximately constant. It is worth mentioning that each subsystem must have a low particle density in order for the Boltzmann statistics to hold for each one of them regardless the temperature.

Thus the system is analyzed as a space-time average of Boltzmann factors $e^{-\tilde{\beta}\hat{H}}$, where $\hat{H}$ is the Hamiltonian of a single subsystem, and the fluctuations are taken into account by a probability distribution $f(\tilde{\beta},\beta)$. In this sense, the formalism can be regarded as the superposition of two statistics: one referring to the Boltzmann factors $e^{-\tilde{\beta}\hat{H}}$ and other to $\tilde{\beta}$.

Mathematically, an averaged Boltzmann factor can be defined as
\bea
\mathcal{B}(\hat{H})=\int_0^\infty f(\tilde{\beta},\beta)e^{-\tilde{\beta}\hat{H}}d\tilde{\beta},
\label{Bdef}
\eea
which leads to the identification of the SE partition function as
\bea
\mathcal{Z}=\text{Tr}\left\{\mathcal{B}(\hat{H})\right\}.
\label{ZSEdef}
\eea 

\begin{figure}
\centering
\includegraphics[scale=.42]{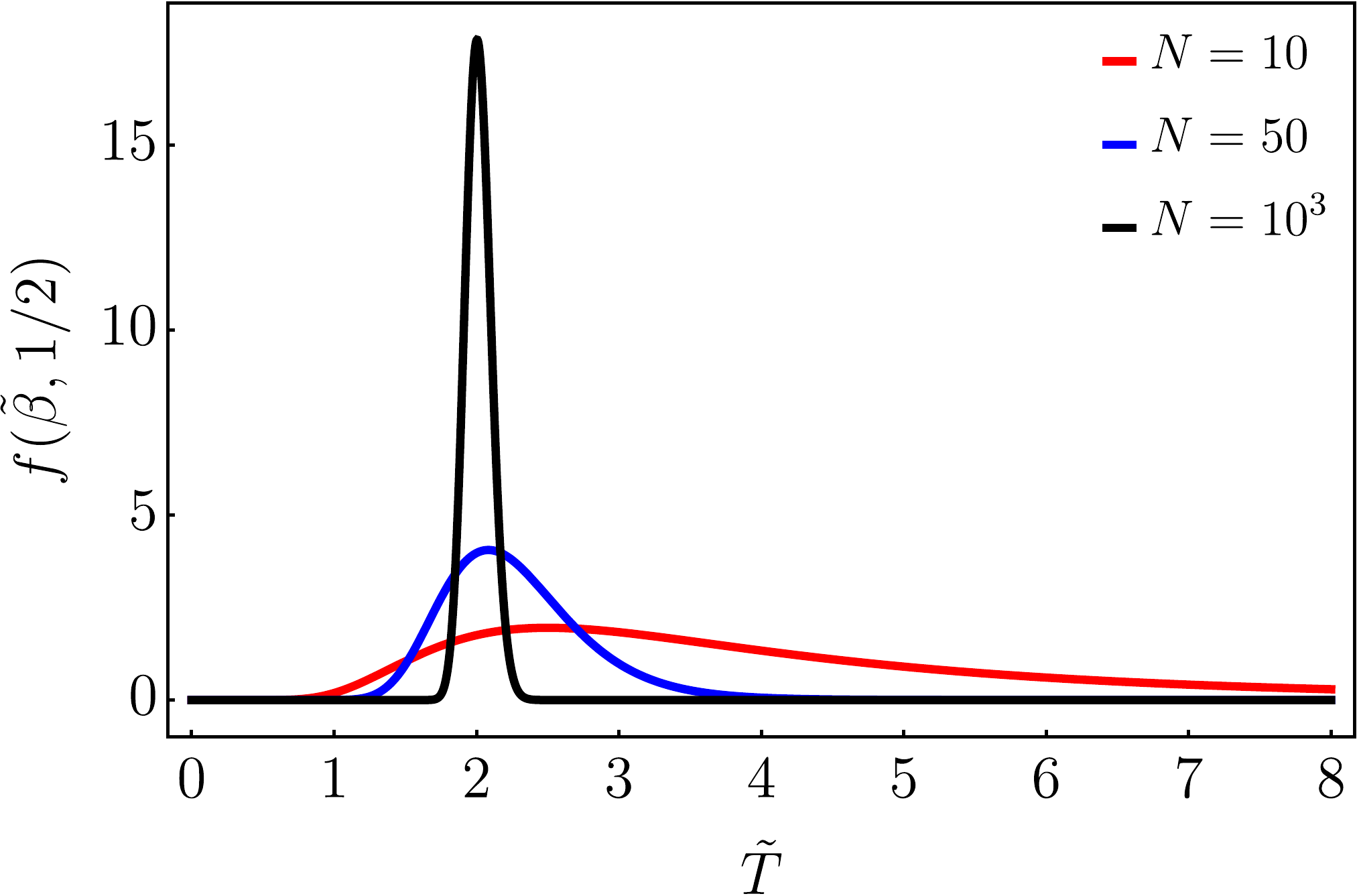}
\caption{$\chi^2$-distribution from Eq.~(\ref{chidist}) as a function of the inverse fluctuating temperature $\Tilde{T}$ (we set $k_B=1$ and $\beta=1/2$) for different values of $N$. For large $N$ the distribution can be approximated as a Dirac delta and the temperature of the whole system is uniquely defined.}
\label{FIG:distfunct}
\end{figure}

Several choices are possible for the distribution obeyed by the parameter $\tilde{\beta}$ over the ensemble of $N$-subsystems, each one of them leading to their corresponding Boltzmann factors~\cite{Obregon1}. In the present document, we chose a $\chi^2$ distribution
\bea
f(\tilde{\beta},\beta)=\frac{1}{\Gamma(N/2)}\left(\frac{N}{2\beta}\right)^{N/2}\tilde{\beta}^{N/2-1}e^{-N\tilde{\beta}/2\beta},
\label{chidist}
\eea
where $\Gamma(x)$ is the gamma function. This distribution has the property that the parameter $\beta$ from Eq.~(\ref{chidist}) corresponds to the average inverse temperature of the whole system, given that 
\bea
\langle\tilde{\beta}\rangle=\int\tilde{\beta}f(\tilde{\beta},\beta)\,d\tilde{\beta}=\beta.
\eea
Because of this, any further thermodynamic observable will refer to the average temperature of the whole system $\beta^{-1}$ and not the individual temperatures $\tilde{\beta}^{-1}$ of the composing subsystems.

By replacing Eq.~(\ref{chidist}) into Eq.~(\ref{Bdef}), the averaged Boltzmann factor corresponding to a $\chi^2$-distribution is given by
\bea
\mathcal{B}(\hat{H})=\left(1+\frac{2}{N}\beta\hat{H}\right)^{-N/2}.
\label{B(H)}
\eea

The $\chi^2$-distribution is a typical distribution for positive-valued random variables. Additionally, if $\chi^2$-like fluctuations evolve on a long timescale, one ends up with Tsallis statistics in a natural way. Tsallis distributions can easily be related to the fact that there are spatio-temporal fluctuations of an intensive parameter such as the inverse temperature.
For other distributions of the intensive parameter, one ends up with more general superstatistics \cite{BeckCohen1} which contain Tsallis statistics as a special case.
For $\hat{H} \ll \beta$, all superstatistics have been shown to have the same first-order corrections to the Boltzmann factor of ordinary statistical mechanics as Tsallis statistics \cite{BeckCohen1}. For moderately large $\hat{H}$, the behaviour of the system is often observed to remain similar to that given by Tsallis statistics \cite{BeckTsalis}. For very large values of $\hat{H}$ the correction to the Boltzmann factor is strongly dependant on the chosen distribution for the fluctuations~\cite{BeckTsalis, Beck2}.  In Subsection~\ref{Subsec:Z} we obtain an exact expression for the SE-partition function independent of the energy scale and the number of subsystems by summing over all the energy spectrum. Because of this, our construction will necessarily lead to Tsallis entropy regardless of the involved parameters and therefore, a proper prescription for computing the thermodynamic observables is strongly required.

\subsection{Partition function and thermodynamic quantities}\label{Subsec:Z}

By taking the trace prescription from Eq.~(\ref{ZSEdef}) of the generalized Boltzmann factor in Eq.~(\ref{B(H)}) for a system with the quantum numbers from Subsection~\ref{subsec:QD}, the SE-partition function turns to be
\bea
\mathcal{Z}_N(\beta)=\sum_{s=\pm1}\sum_{n=0}^{+\infty}\sum_{l=-\infty}
^{+\infty}\left(1+\frac{2}{N}\beta E_{nls}\right)^{-N/2}.
\label{Zprevia}
\eea
The sum over the energy levels from Eq.~(\ref{energyspectrum}) can be analytically found to be
\bea
\mathcal{Z}_N(\beta)&=&\sum_{s=\pm1}\left\{\Lambda^{(-)}_s(\beta,N)\sum_{n=0}^{\infty}\zeta\left[\frac{N}{2},\Delta_{ns}^{(-)}(\beta,N)\right]\right.\nn\\
&+&\left.\Lambda^{(+)}_s(\beta,N)\sum_{n=0}^{\infty}\zeta\left[\frac{N}{2},1+\Delta_{ns}^{(+)}(\beta,N)\right]\right\},\nn\\
\eea
where $\zeta(a,x)$ is the Hurwitz zeta function and the auxiliary functions $\Lambda^{(\pm)}_s$ and $\Delta_{ns}^{(\pm)}$ are defined as
\begin{subequations}
\bea
\Lambda^{(\pm)}_s(\beta,N)=\left[\frac{N}{2\beta\hbar\left(\Omega_s\pm\omega_c/2-\gamma m^*\omega^2s/\hbar\right)}\right]^{N/2}
\eea
and
\bea
\Delta_{ns}^{(\pm)}(\beta,N)=\frac{N/2\beta\hbar+(2n+1)\Omega_s+g^*\omega_c s/4-V_0/\hbar}{\Omega_s\pm\omega_c/2-\gamma m^*\omega^2s/\hbar}.\nn\\
\eea
\end{subequations}

Certainly, the partition function contains the full thermodynamic of the system which can be reached from the derivatives of its natural logarithm. 
To achieve an accurate description of the system's response functions, note that the Eq.~(\ref{Zprevia})  corresponds precisely with the partition $Z_q$ of the Tsallis probability distribution~\cite{TsallisZ2}
\begin{subequations}
\bea
p_n(\beta)=\frac{\left[1-(1-q)\beta E_n\right]^{1/(1-q)}}{Z_q(\beta)},
\label{pn}
\eea
with
\bea
Z_q(\beta)\equiv \sum_n\left[1-(1-q)\beta E_n\right]^{1/(1-q)}.
\label{Zq}
\eea
\end{subequations}
The entropic index $q=1+2/N$ characterizes the degree of subextensivity of the entropy addition rule as the usual Boltzmann entropy is recovered in the limit $q\rightarrow 1$ or $N\rightarrow\infty$.

\begin{figure*}
\centering
\includegraphics[scale=.47]{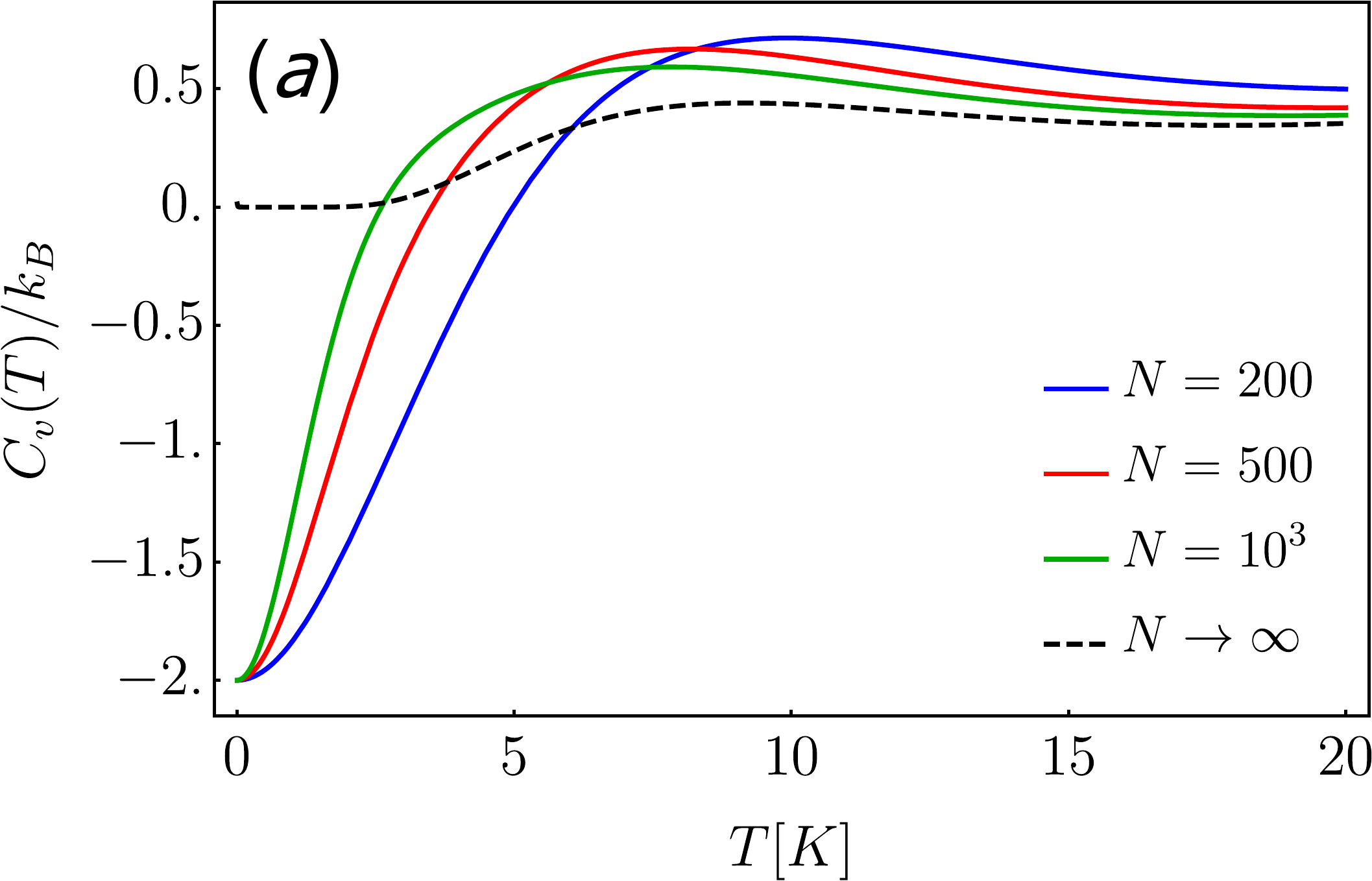}
\hspace{0.5cm}
\includegraphics[scale=.6]{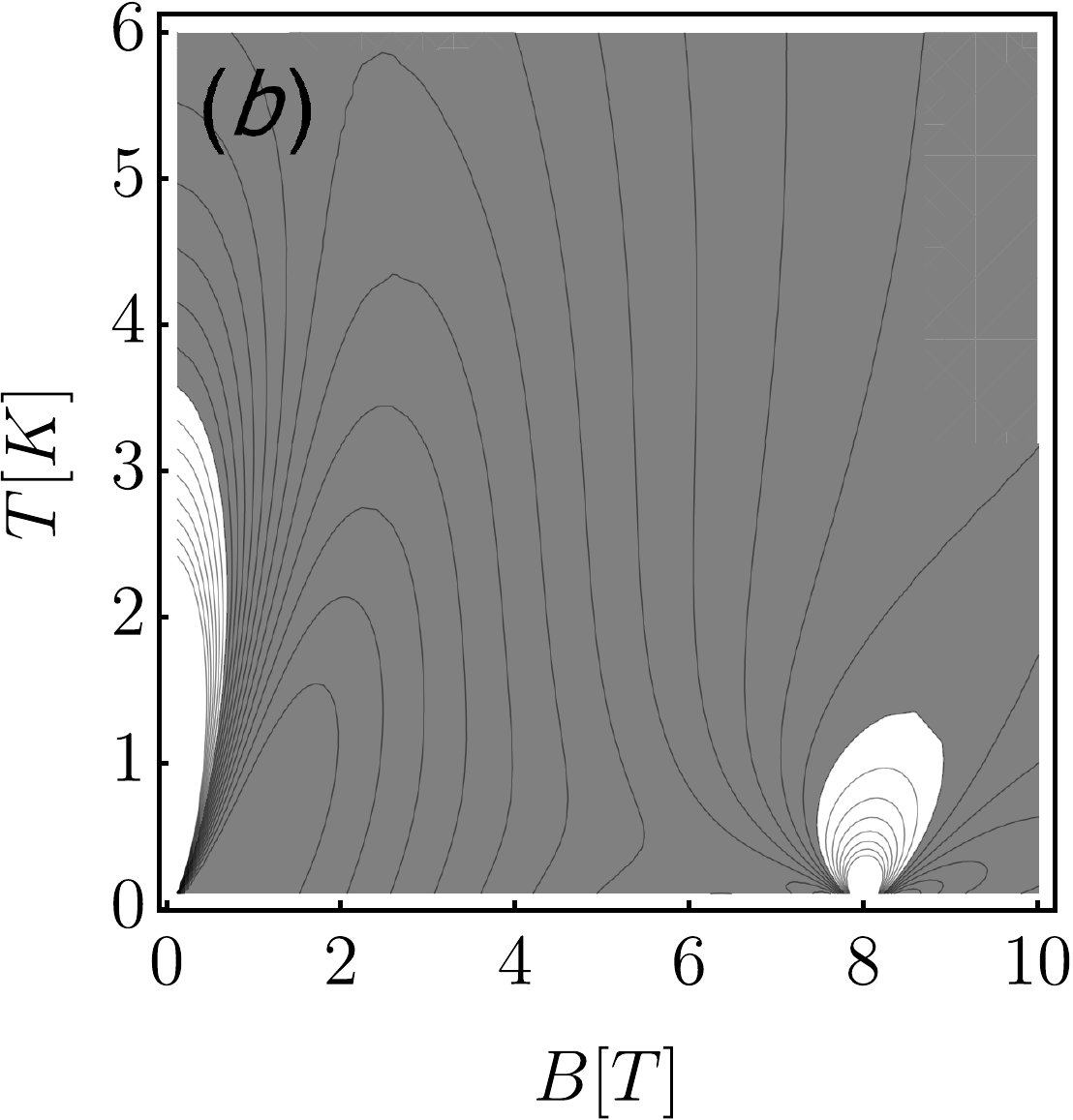}
\caption{  Thermodynamic functions by following the natural logarithm prescription of Eqs.~(\ref{Cvandchiln}) for the SE-partition function. Panel (a) displays the numerical calculations of the specific heat as a function of temperature using a Rashba spin-orbit coupling of $\gamma=0.15$ nm$^2$ for $B=5$~T and $V_0= 20$ meV. The panel \textbf{(b)} shows the magnetic phase diagram for $V_0=8$ meV, $N=100$, and  $\gamma=0.15$ nm$^2$. The gray region corresponds to the
diamagnetic phase ($\chi<0$), whereas the white region to the
paramagnetic phase ($\chi>0$).  } \label{FIG:CvandChiZ1}
\end{figure*}

Tsallis \etal~\cite{TsallisZ2} showed that in order to preserve the Legendre structure of the statistical thermodynamics, the $q$-logarithm has to be introduced~\cite{Tsallis_q-log}: 
\bea
\ln_qx\equiv\frac{x^{1-q}-1}{1-q},\;\;\text{so that}\;\;\ln_1 x=\ln x,
\label{lnq}
\eea
and therefore, the Free Energy, the average energy and the specific heat are given by
\begin{subequations}
\bea
F_q(\beta)\equiv U_q(\beta)-TS_q=-\frac{1}{\beta}\ln_qZ_q(\beta),
\eea
\bea
U_q(\beta)=-\frac{\partial}{\partial\beta}\ln_qZ_q(\beta)
\eea
and
\bea
C_q(\beta)=\frac{\partial U_q(\beta)}{\partial T}.
\label{CvZ2}
\eea
\label{LegendreStructure}
\end{subequations}
In the expression above $S_q$ is the well-know non-extensive Tsallis entropy~\cite{Tsallis_original}:
\bea
S_q=k_B\frac{1}{q-1}\left(1-\sum_n p_n^q\right)\forall\;q \in\mathbb{R},
\label{TsallisEntropy}
\eea
with the constriction
\bea
\sum_n p_n=1.
\eea

The use of the natural logarithm instead of the $q$-logarithm given in Eq.~(\ref{lnq}) will therefore break the Legendre structure of the potentials.
Even though it could be argued that there is no reason to believe that in a non-extensive scenario the Legendre structure has to remain valid, it must be stressed that in such scenario the parameter $\beta$ in Eqs.~(\ref{LegendreStructure}) is not the Lagrange multiplier associated to the internal energy constraint. Furthermore, as studied in the field of Geometrothermodynamics (GTD), Legendre transformations can be regarded as diffeomorphisms that leave the space of equilibrium states unchanged. In such manner, the curvature of the induced metric in the space of equilibrium states is truly independent of the thermodynamic potential used to describe a given system~\cite{GTD-Alessandro}. Because of this, a stability criterion within the Tsallis formalism is well-defined via the positivity of the specific heat~\cite{NegHeat_stability}. 

Additionally, the $q$-logarithm corresponds to the definition of the $q$-expectation values as~\cite{Tsallis_q-log,TsallisZ2}
\bea
\langle \mathcal{O} \rangle_q =\sum_n p_n^q(\beta)\mathcal{O}_n.
\eea
In particular, from the $q$-expectation value for the internal energy 
\bea
U=\sum_n p_n^q(\beta)E_n,
\eea
the parameter $\beta$ can be associated with the Lagrange parameter of the average energy and the Legendre structure is naturally recovered. 

In~\cite{TsallisZ2}, a prescription to construct a \textit{renormalized temperature} from the parameter $\beta$  is given in order to avoid some undesirable features of the theory such as non-additivity of the internal energy and loss of norm $\langle 1 \rangle_q \ne 1$. In the present work we are not evaluating the effects of such a renormalization but we emphasize that the use of the $q$-logarithm already reproduces most of the features that a thermodynamic theory requires.

\begin{figure*}
\centering
\includegraphics[scale=.38]{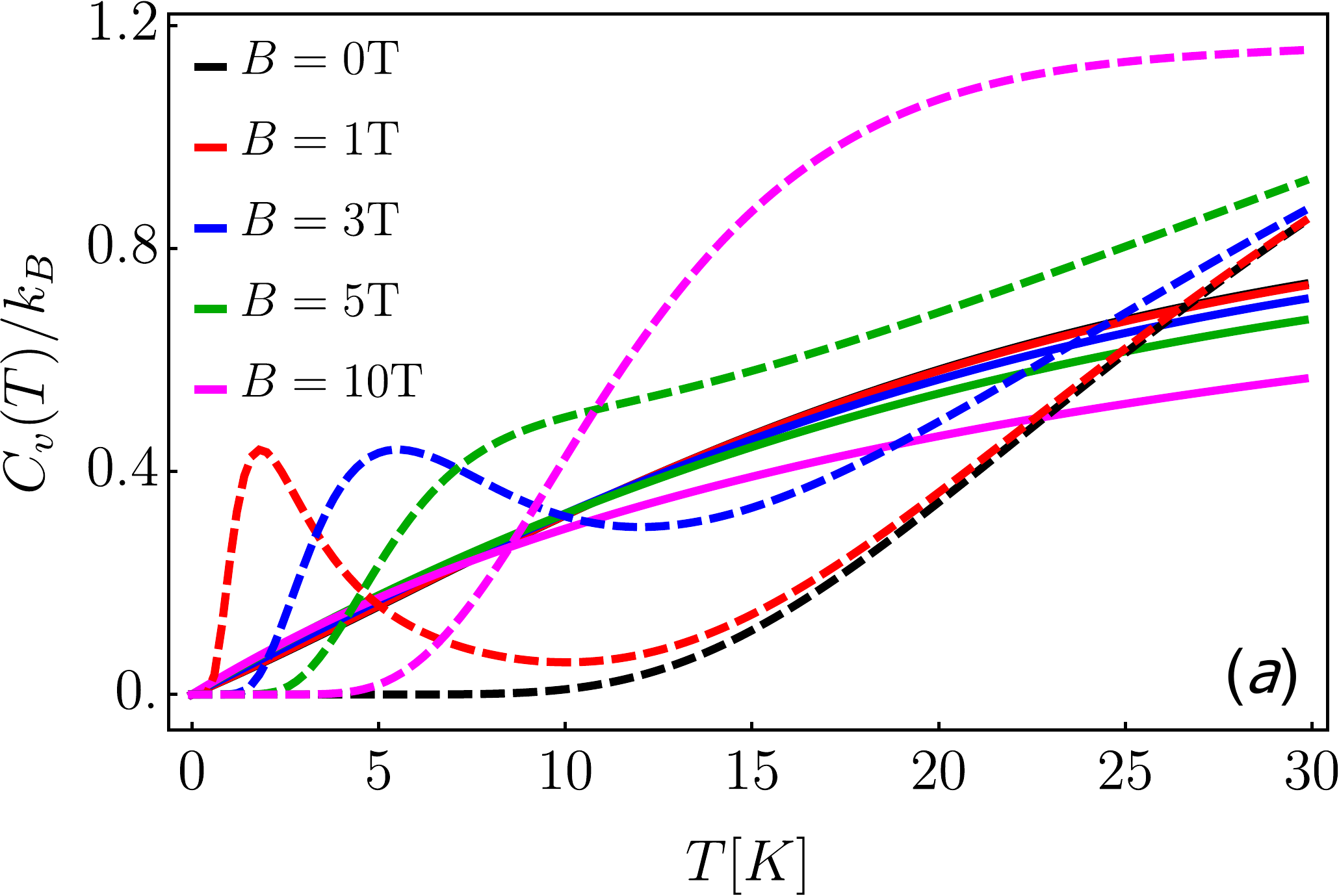}\hspace{0.4cm}\includegraphics[scale=.38]{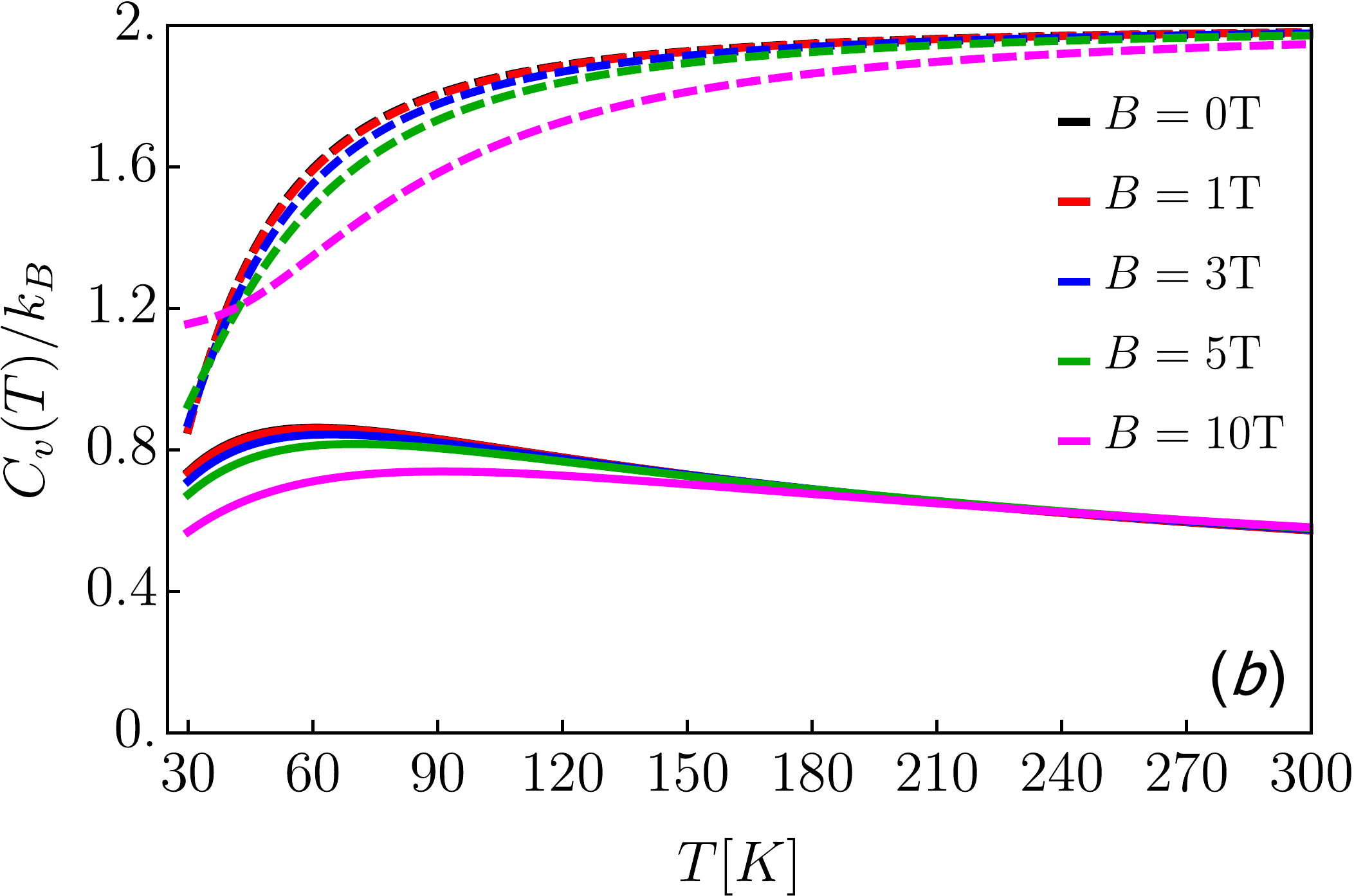}\\
\vspace{0.3cm}
\includegraphics[scale=.38]{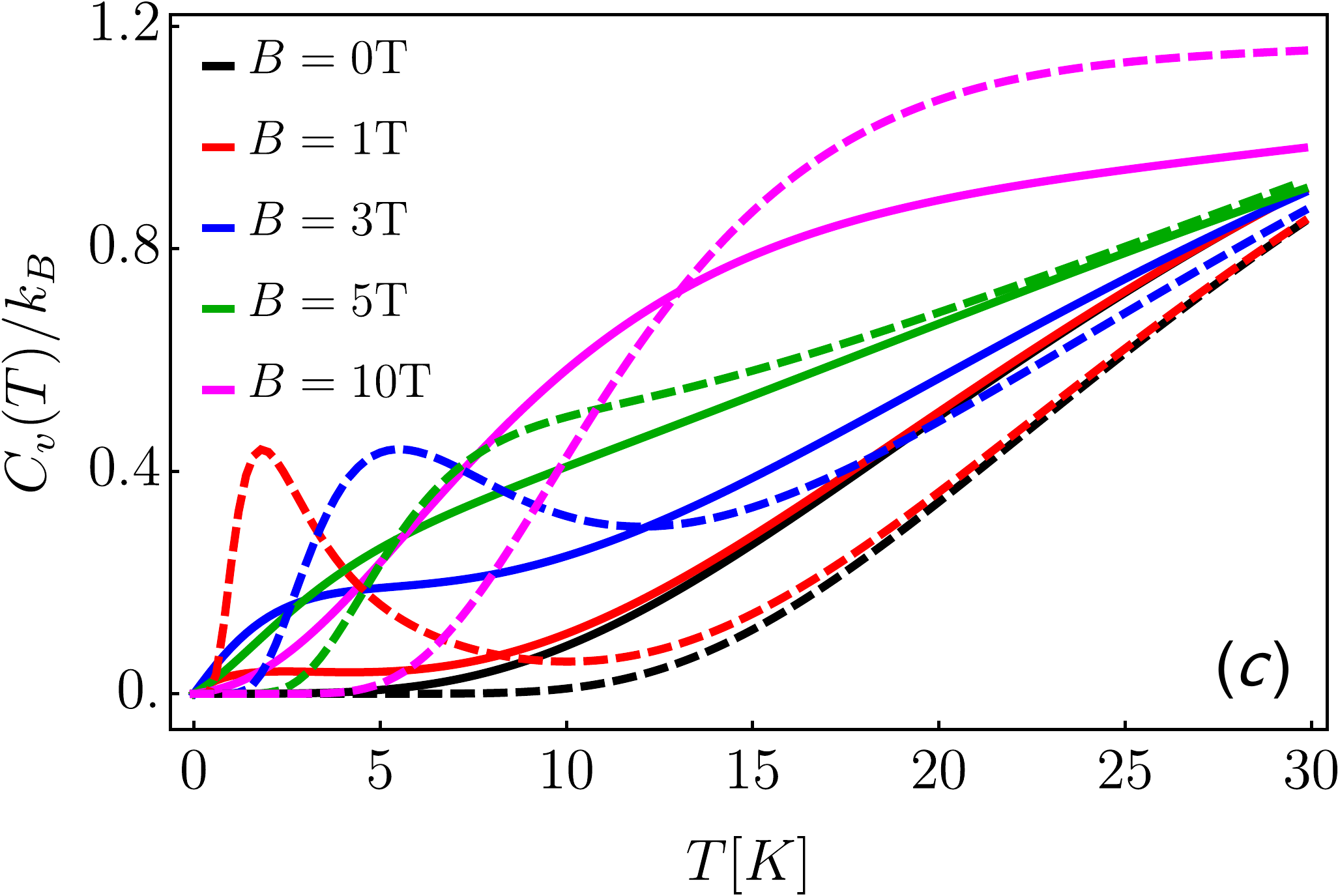}\hspace{0.4cm}\includegraphics[scale=.38]{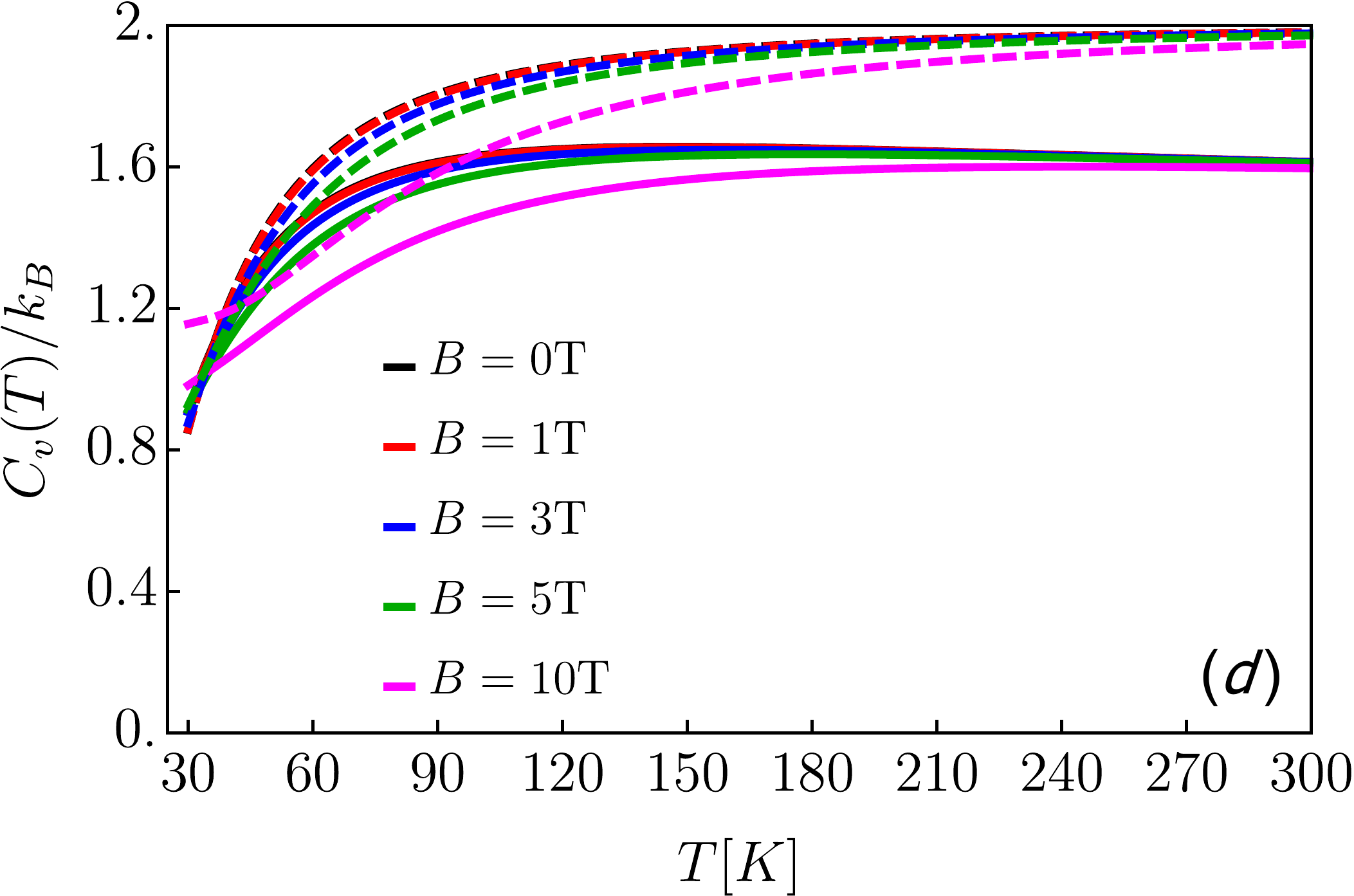}\\
\vspace{0.3cm}
\includegraphics[scale=.38]{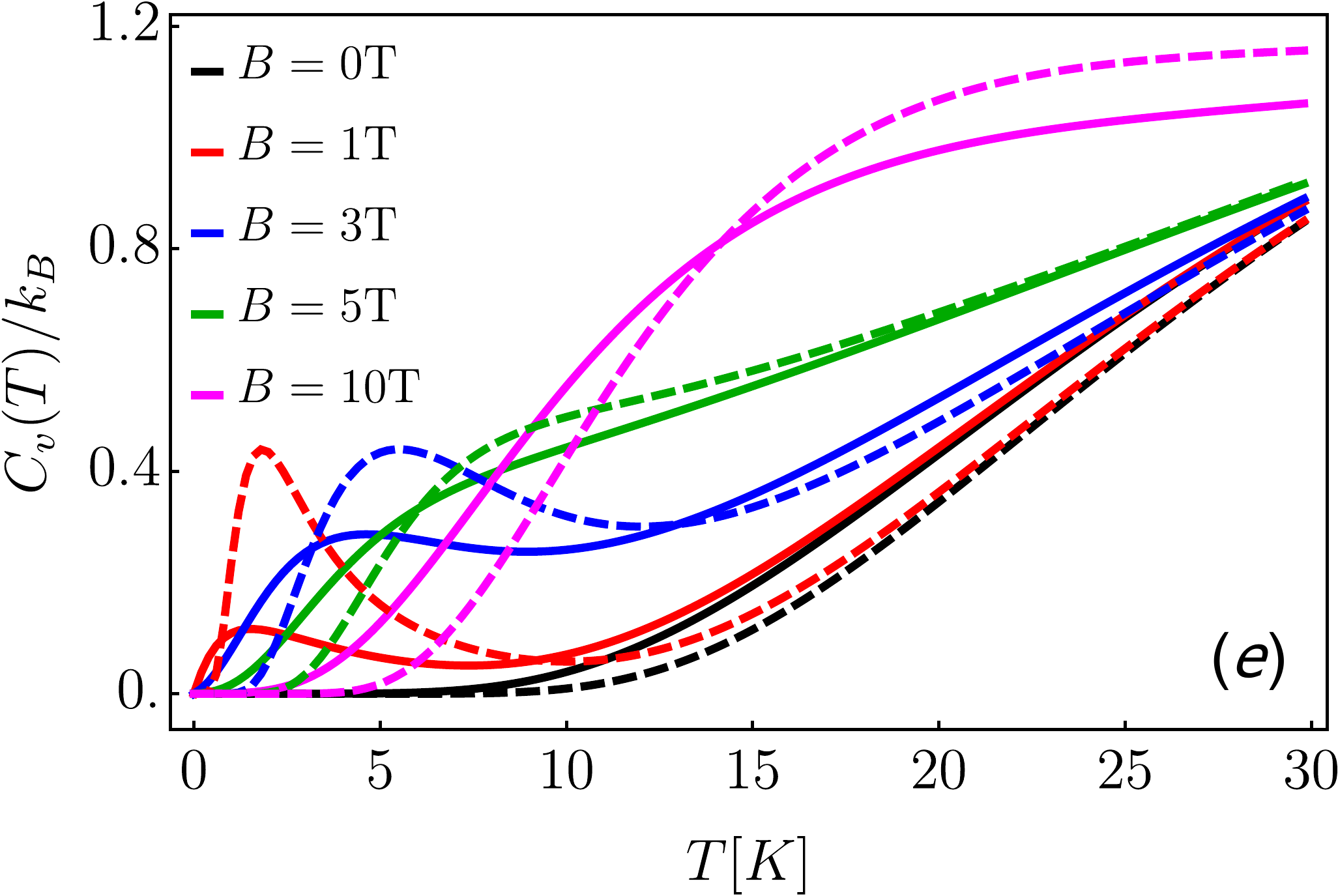}\hspace{0.4cm}\includegraphics[scale=.38]{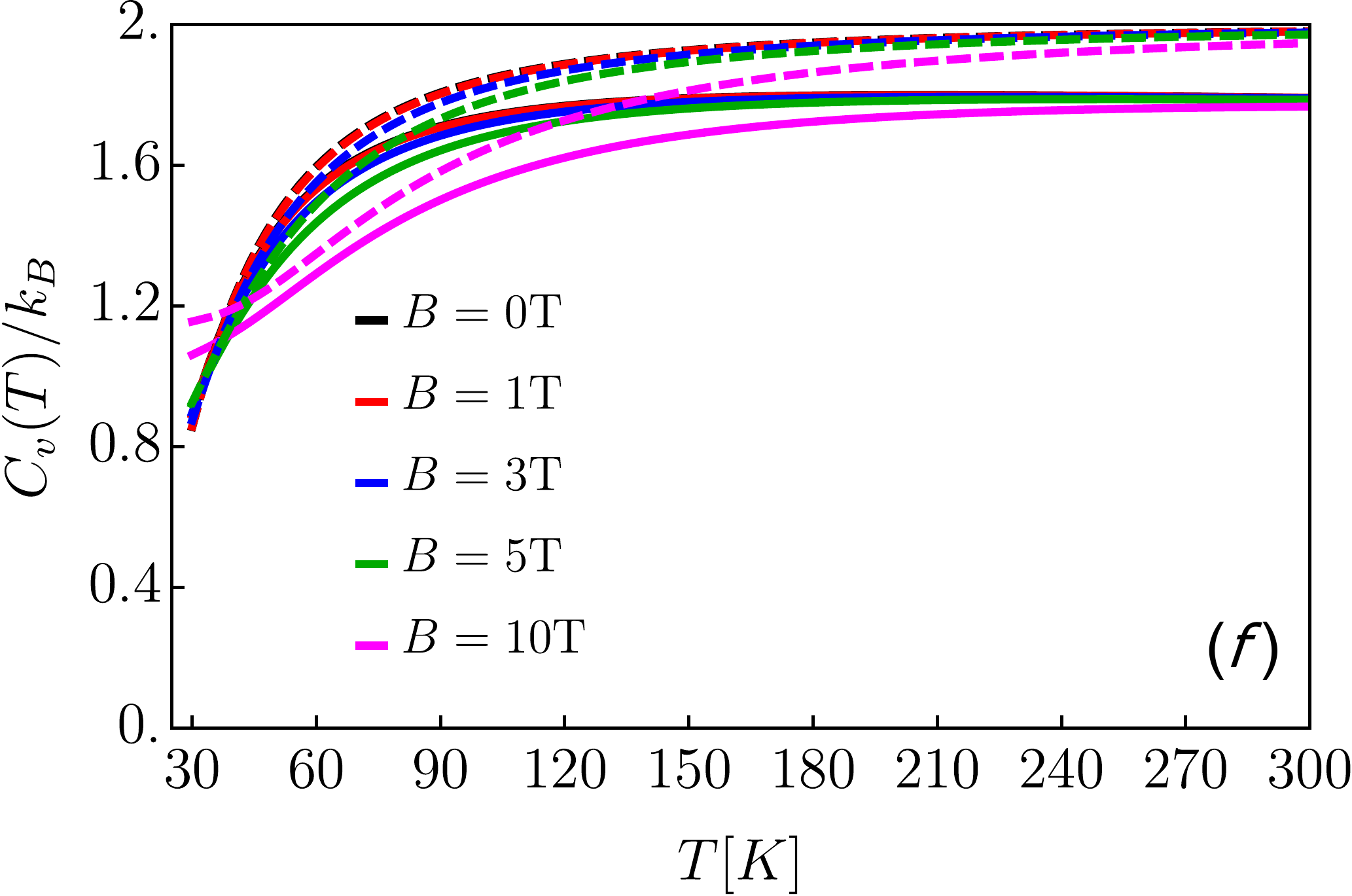}\\
\vspace{0.3cm}
\includegraphics[scale=.38]{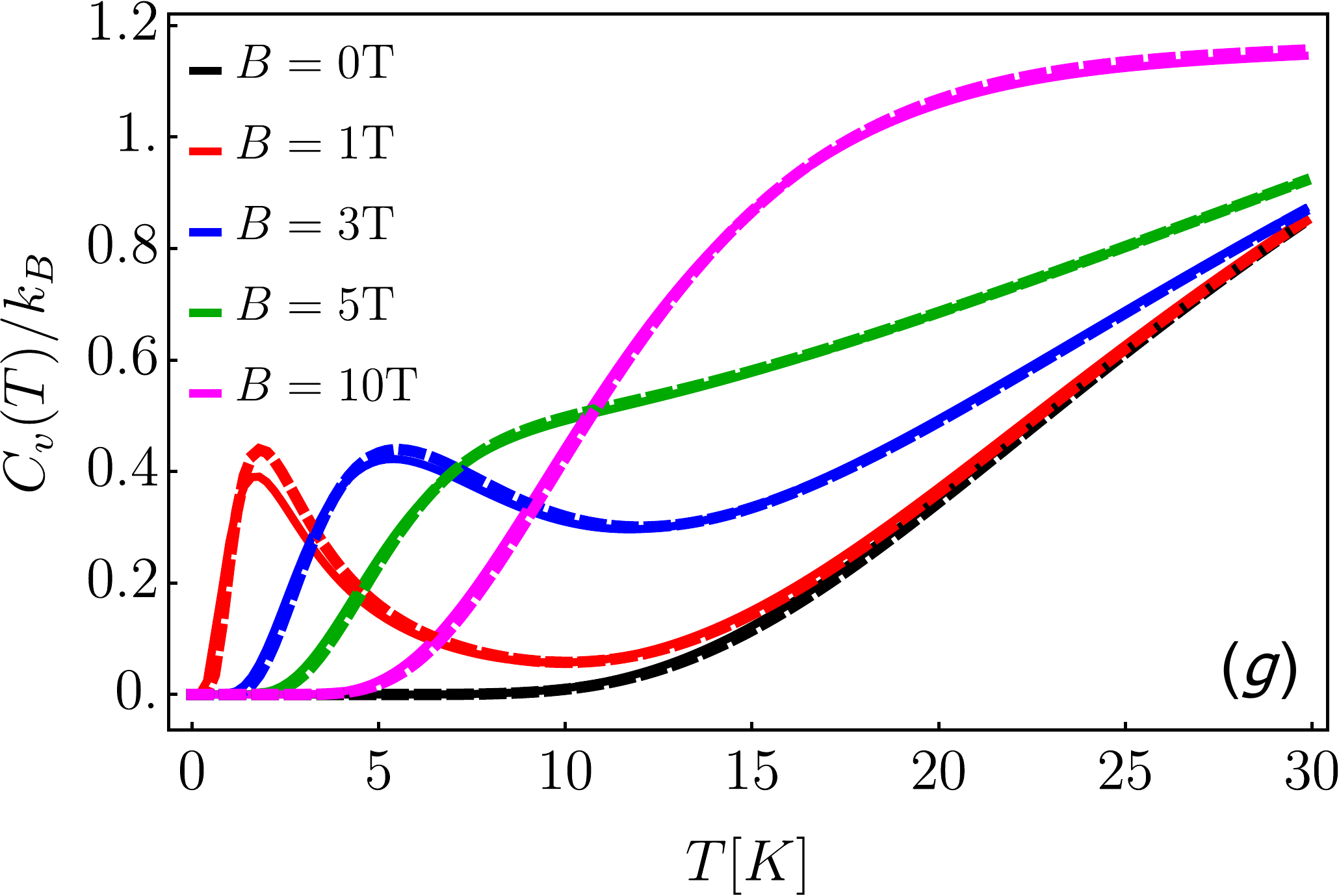}\hspace{0.4cm}\includegraphics[scale=.38]{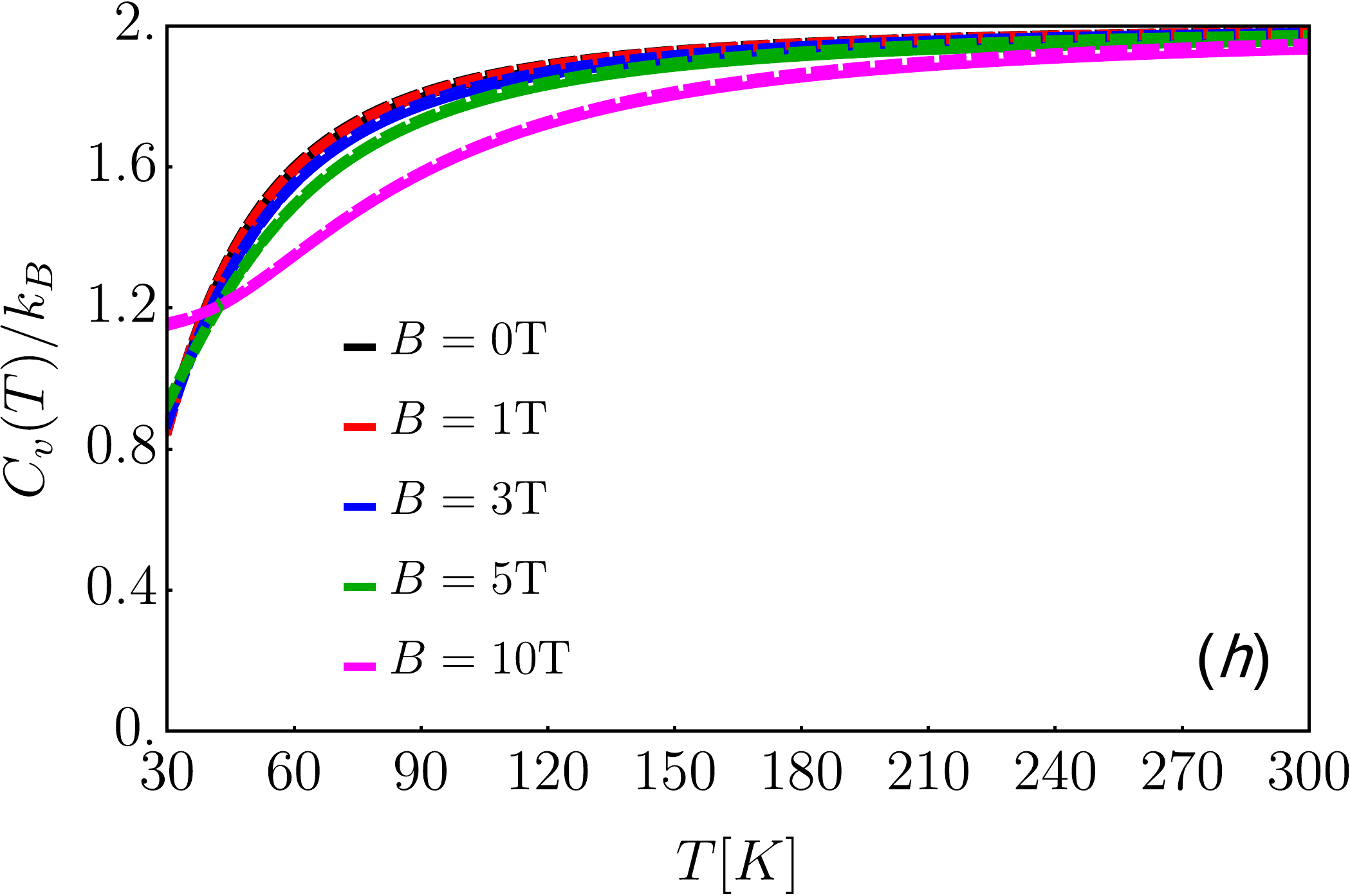}
\caption{  Specific heat as a function of temperature using a Rashba spin-orbit coupling of $\gamma=0.15$ nm$^2$ for increasing values of the external magnetic field $B$ and different system sizes: $N=10$ for panels (a) and (b); $N=50$ for panels (c) and (d); $N=100$ for panels (e) and (f) and $N=10^3$ for panels (g) and (h). In each panel, the solid lines represent the super-statistical behaviour obtained from Eq.~(\ref{CvZ2}), while the Boltzmann statistics are depicted as doted lines and can regarded as the asymptotic behaviour for $N\rightarrow\infty$.  }
\label{FIG:CvconZ2}
\end{figure*}

\section{Results and discussion}
\label{sec:results}

In order to compute the thermal and magnetic properties of our system, we set the parameters as follows: $m^{*}$=0.067$m_0$ is the effective electron mass for a GaAs QD where $m_0$ is the free electron mass; $g^{*}=-0.44$ is the effective
Land\'e constant, $\gamma$=0.15 nm$^2$, $R$=10 nm is the radio of the QD, and $V_0=10$ meV as the value of the confining potential.\\

In Fig.~\ref{FIG:CvandChiZ1} we show (a) the specific heat $C_v$ and (b) the magnetic susceptibility $\chi$ obtained by taking the natural logarithm of the SE-partition function, \textit{i.e.}, by using the relations
\bea
C_v&=&k_B\frac{\partial}{\partial T}\left(T^2\frac{\partial}{\partial T}\ln\mathcal{Z}_N\right),\nn\\ \chi&=&k_BT\frac{\partial^2}{\partial B^2}\ln\mathcal{Z}_N,
\label{Cvandchiln}
\eea 
This will lead to nonphysical results such as a negative specific heat in the low temperature region and the emergence of a spurious paramagnetic region at high external magnetic fields. Both phenomena are persistent even for a large number of subsystems. This is an effect of the truncation scheme of the Taylor series of the natural logarithm in comparison with the $q$-logarithm, namely
\bea
\ln_q \mathcal{Z}_N &=&
\ln \mathcal{Z}_N - \frac{1}{N}\ln^2 \mathcal{Z}_N
+ \mathcal{O}(N^{-2}) \nn\\
&\approx& 
\ln\mathcal{Z}_0+\frac{1}{N}\left(\frac{\beta^2}{\mathcal{Z}_0}\frac{\partial^2\mathcal{Z}_0}{\partial\beta}-\ln^2\mathcal{Z}_0\right) ,
\label{log-corrected}
\eea
 where $\mathcal{Z}_0$ is the partition function coming from the standard formalism of physical statistics. This means that for large but finite number of subsystems $N$, the out-of-equilibrium effects will be enhanced in the low-temperature regime. Moreover, the introduction of the $q$-logarithm will lead to corrections to the response functions of $\mathcal{O}\left( \frac{1}{N}\ln\mathcal{Z}_0 \right)$ that will only be negligible when the natural logarithm of the partition function itself is small with respect to the number of subsystems $N$. This will in general depend in a non-trivial way on the thermodynamic variables of $\mathcal{Z}_0 $.\\

From Fig.~\ref{FIG:CvandChiZ1}-(a) it can be noticed that for some values of $N$ and $T$, the specific heat becomes negative and always tends to $-2k_B$ as $T\rightarrow0$. Adding the correction from Eq.~(\ref{log-corrected}) will recover the $N\rightarrow\infty$ behaviour depicted in the dotted line and positivity is restored.
It is worth mentioning that if either the interactions are long-range or if the system is small concerning the interaction range, then negative specific heats can be, and indeed have been, observed~\cite{NegHeat_condition}. Such systems include among others gravitational systems (long-range interactions) \cite{NegHeat_gravity1, NegHeat_gravity2, NegHeat_gravity3} and atomic clusters (small systems) \cite{NegHeat_atomic1, NegHeat_atomic2, NegHeat_atomic3}. Nevertheless, we attribute such behaviour in the system under study in the present work to a poor choice of the definition of the prescription for obtaining physical observables.

On the other hand, by comparison with Refs.~\cite{packman} and~\cite{Castano}, the magnetic phase diagram from Fig.~\ref{FIG:CvandChiZ1}-(b) shows the emergence of a paramagnetic phase (represented by the white region at high magnetic fields). Those results are non-intuitive and contradict the stability condition for the subsystems conforming the array of QD's~\cite{NegHeat_definition}. By this means, the necessity for introducing $q$-expectation values becomes more evident.

 It is worth mentioning that several authors have used the standard definition of an expectation value for their calculations in order to explore the thermodynamic consequences of a superstatistical treatment in different scenarios: the thermodynamical properties of the anharmonic canonical ensemble within the cosmic-string framework~\cite{Sobhani1}, the impact of the non-commutativity of the space for systems with thermal fluctuations~\cite{Sargolzaeipor2}, the effective Quantum Chromodynamics phase diagram~\cite{AyalaCEP} and the anharmonic oscillator for non-relativistic and relativistic Klein-Gordon equations~\cite{WangKG}, among others~\cite{Sargolzaeipor4,SSQD}. We suggest the restoration of the Legendre invariance in Refs.~\cite{Sargolzaeipor2,Sobhani1,WangKG,AyalaCEP,SSQD,Sargolzaeipor4} to regain a closed thermodynamic treatment. 

In the following, only $q$-expectation values obtained from the full SE-partition function are evaluated and discussed.\\

The specific heat $C_v$ as a function of temperature for several values of the external magnetic field is shown in Fig.~\ref{FIG:CvconZ2}. To appreciate the effect of SE, we compare the results for a varying number of subsystems $N$ with respect to the extensive Boltzmann statistics represented by the respective dotted lines. As can be noticed from Figs.~\ref{FIG:CvconZ2} (a), (c), (e) and (g), the Schottky anomaly slowly disappears as the full system decreases in size. On the other hand, Figs.~\ref{FIG:CvconZ2} (b), (d), (f) and (h) show that for a small number of subsystems  the specific heat ceases to be a monotonically increasing function with respect to the average temperature. 

This behavior can be understood in terms of the probability distribution function of Eq.~(\ref{chidist}) which is plotted if Fig.~\ref{FIG:distfunct} for $\beta=1/2$ ($T=2$ when $k_B=1$) and for different values of $N$. For a small number of subsystems and a small average temperature, a wide range of fluctuations can contribute to the weighted Boltzmann factor. 
In contrast, the distribution function for a very large number of composing systems $N\rightarrow\infty$ resembles a Dirac delta and the temperature of the whole system can be approximated as a unique value and thermodynamic equilibrium is established. 
Nevertheless, when the number of subsystems is large but finite, the $\chi^2$-distribution will get narrower for increasing values of the average temperature, asymptotically approaching to a Dirac delta for the infinite temperature limit. Therefore, one can expect high deviations from the Boltzmann statistics for systems near room average temperature whenever they are far from the thermodynamic equilibrium. 

The low-temperature peak of the Schottky anomaly is closely related to the energy required for a thermal transition between the ground state and the first excited state of the system (with energy $\Delta E$) as it can be interpreted as a resonance in $k_B T_s \sim \Delta E$. In the case of a broad distribution of temperatures (small values of $N$) the resonance also becomes broader and eventually smears out.

The specific heat in the high-temperature regime asymptotically approaches a constant as would be expected from a Dulong-Petit-like behaviour. Surprisingly, this is truth even for relatively small number of subsystems, where the value for such asymptotic constant seems to be lower for decreasing values of $N$. In other words, out-of-equilibrium effects will introduce corrections that effectively lower the heat capacity of the system. This seems to be a consequence of the $\chi^2$-distribution with low number of subsystems, where contributions from $\tilde{\beta}$ lower than the average $\beta$ (i.e., at high temperatures) are dominant, meaning that most of the subsystems already have higher local temperatures than the average temperature of the system and therefore less heat transfer is necessary to increase it. \\

\begin{figure}
    \centering
    \includegraphics[scale=0.4]{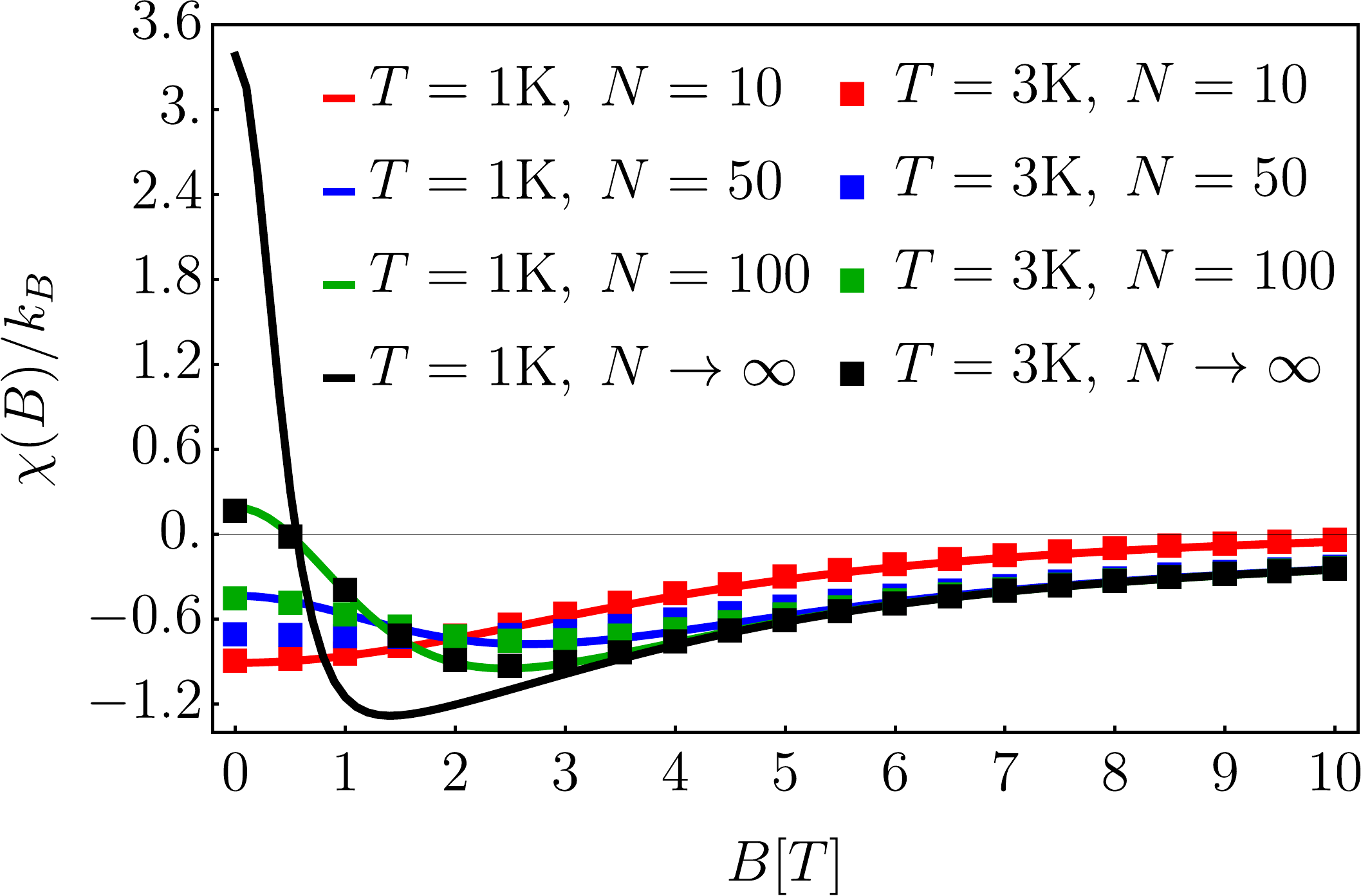}
    \caption{ $q$-expectation value for the magnetic susceptibility as a function of the external magnetic field using a Rashba spin-orbit coupling of $\gamma=0.15$ nm$^2$ for for different values of temperature and number of composing subsystems as labeled in the Figure. The solid black line and black dots are the asymptotic behaviour for $N\rightarrow\infty$. } 
    \label{fig:susceptibility}
\end{figure}

 Finally, it is noteworthy from Fig.~\ref{fig:susceptibility} that the paramagnetic region in the low-temperature low-external field regime typical from GaAs QDs \cite{packman,Castano} disappears when the out-of-equilibrium corrections are introduced using the $\chi^2$-SE formalism for decreasing number of subsystems $N$. This is physically interpreted as a lack of order in the spins coming from the fact that for small $N$, a low average temperature will still have important contributions from subsystems with higher local temperature that will break the order of the spins in a global description.

\section{Summary}\label{sec:conclusions}

 We examined the effect of non-equilibrium processes modeled by the $\chi^2$-superstatistics on the thermal and magnetic properties of an array of two-dimensional GaAs quantum dots with Rashba spin-orbit interaction in the presence of an external uniform and constant magnetic field. 

First, we used the present model to quantitatively emphasize the importance of an appropriate construction of physical observables for obtaining a correct description of the physics derived from a non-extensive construction of the entropy.

Afterwards, we offered an improved calculation obtained from the analytic solution for the partition function. This allowed us to study the impact of an arbitrary number of subsystems on the superstatistical corrections and confirms that the ordinary thermo-magnetic properties are recovered whenever the thermal distribution of the composing subsystems can be approximated by a Dirac delta. 

Finally, we found that the most remarkable out-of -equilibrium effects appear for a small number of subsystems or at the low-temperature regime, this is, whenever the $\chi^2$-distribution is spread over a large range of temperatures. In terms of the response functions, this means that the introduction of a broad range of fluctuations in the local temperatures of the system is responsible for a progressive disappearance of the Schottky anomaly, while the high-average temperature specific heat gets effectively decreased. Furthermore, a small number of composing subsystems is found to suppress the paramagnetic phase transition that would otherwise appear at low temperatures.

%

\section{ACKNOWLEDGMENT}
The authors J. D. Castaño-Yepes and D. A. Amor-Quiroz acknowledge financial support from Consejo Nacional de Ciencia y Tecnolog\'ia (CONACyT). The authors also thank Dr. Alessandro Bravetti for useful comments about the importance of the Legendre structure of a thermodynamic theory.
\\\\

\end{document}